\documentclass[fleqn,twoside]{article}
\usepackage{espcrc2}

\usepackage{graphicx}
\usepackage[figuresright]{rotating}

\def\la{\mathrel{\mathpalette\fun <}}
\def\ga{\mathrel{\mathpalette\fun >}}
\def\fun#1#2{\lower3.6pt\vbox{\baselineskip0pt\lineskip.9pt
  \ialign{$\mathsurround=0pt#1\hfil##\hfil$\crcr#2\crcr\sim\crcr}}}

\newcommand{\AmS}{{\protect\the\textfont2
  A\kern-.1667em\lower.5ex\hbox{M}\kern-.125emS}}

\title{Cosmic Magnetic Fields and Their Influence on Ultra-High
Energy Cosmic Ray Propagation}

\author{G\"unter Sigl\address{
GReCO, Institut d'Astrophysique de Paris, C.N.R.S.,
98 bis boulevard Arago, F-75014 Paris, France\\
F\'{e}d\'{e}ration de Recherche Astroparticule et Cosmologie, 
Universit\'{e} Paris 7, 2 place Jussieu, 75251 Paris Cedex 05, France}
Francesco Miniati, Torsten~A.~En\ss lin\address{
Max-Planck Institut f\"ur Astrophysik,
Karl-Schwarzschild-Str.~1, 85741 Garching, Germany}}
       
\begin{document}

\begin{abstract}
We discuss the influence of large scale cosmic magnetic fields
on the propagation of hadronic cosmic rays above $10^{19}\,$eV
based on large scale structure simulations. Our simulations suggest
that rather substantial deflection up to several tens of degrees at
$10^{20}\,$eV are possible for nucleon primaries. Further, spectra
and composition of cosmic rays from individual sources can depend on magnetic
fields surrounding these sources in intrinsically unpredictable
ways. This is true even if deflection from such individual sources
is small. We conclude that the influence of large scale cosmic magnetic
fields on ultra-high energy cosmic ray propagation is currently
hard to quantify. We discuss possible reasons for discrepant
results of simulations by Dolag et al. which predict deflections
of at most a few degrees for nucleons. We finally point out that even
in these latter simulations a possible heavy component would in general
suffer substantial deflection.

\vspace{1pc}
\end{abstract}

\maketitle

\section{Introduction}
The origin of ultra-high energy cosmic rays (UHECR) above $10^{19}\,$eV
($=10\,$EeV) has been challenging physicists for 
many years~\cite{reviews,school}. Several
next-generation experiments, most notably the Pierre Auger
experiment now under construction~\cite{auger} and the EUSO
project~\cite{euso} are now trying to solve this mystery.

Although statistically meaningful information about the UHECR energy
spectrum and arrival direction distribution has been accumulated, no
conclusive picture for the nature and distribution of the sources
emerges naturally from the data. There is on the one hand the approximate
isotropic arrival direction distribution~\cite{bm} which indicates that we are
observing a large number of weak or distant sources. On the other hand,
there are also indications which point more towards a small number of
local and therefore bright sources, especially at the highest energies:
First, the AGASA ground array claims statistically significant multi-plets of
events from the same directions within a few degrees~\cite{teshima1,bm},
although this is controversial~\cite{fw} and has not been seen so far
by the fluorescence experiment HiRes~\cite{finley}.
The spectrum of this clustered component is $\propto E^{-1.8}$ and thus
much harder than the total spectrum~\cite{teshima1}.
Second, nucleons above $\simeq70\,$EeV suffer heavy energy losses due to
photo-pion production on the cosmic microwave background
--- the Greisen-Zatsepin-Kuzmin (GZK) effect~\cite{gzk} ---
which limits the distance to possible sources to less than
$\simeq100\,$Mpc~\cite{stecker}. Heavy nuclei at these energies
are photo-disintegrated in the cosmic microwave background within a
few Mpc~\cite{heavy}. For a uniform source distribution
this would predict a ``GZK cutoff'', a drop in the spectrum.
However, the existence of this ``cutoff'' is not established yet
from the observations~\cite{bergman}.

The picture is further complicated by the likely presence of large
scale extra-galactic magnetic fields (EGMF) that will lead to deflection
of any charged UHECR component.
Magnetic fields are ubiquitous in the Universe, although their
origin is still unclear~\cite{bt_review}. Magnetic fields
in galaxies are observed with typical strengths of a few
micro Gauss. In addition
there is some evidence for fields correlated
with larger structures such as galaxy clusters~\cite{bo_review}.
Magnetic fields as strong as
$\simeq 1 \mu G$ in sheets and filaments of the large scale galaxy
distribution, such as in our Local Supercluster, are compatible with
existing upper limits on Faraday rotation~\cite{bo_review,ryu}.
It is also possible that fossil cocoons of former radio galaxies,
so called radio ghosts, contribute significantly to the isotropization
of UHECR arrival directions~\cite{mte}.

To get an impression of typical deflection angles one can characterize the
EGMF by its r.m.s. strength $B$ and a coherence length $l_c$.
If we neglect energy loss processes for the moment, then
the r.m.s. deflection angle over a distance $r\ga l_c$ in such a field
is $\theta(E,r)\simeq(2rl_c/9)^{1/2}/r_L$~\cite{wm}, where the Larmor
radius of a particle of charge $Ze$ and energy $E$ is
$r_L\simeq E/(ZeB)$. In numbers this reads
\begin{eqnarray}
  \theta(E,r)&\simeq&0.8^\circ\,
  Z\left(\frac{E}{10^{20}\,{\rm eV}}\right)^{-1}
  \left(\frac{r}{10\,{\rm Mpc}}\right)^{1/2}\nonumber\\
  &&\times\left(\frac{l_c}{1\,{\rm Mpc}}\right)^{1/2}
  \left(\frac{B}{10^{-9}\,{\rm G}}\right)\,,\label{deflec}
\end{eqnarray}
for $r\ga l_c$. This expression makes it immediately obvious
that fields of fractions of micro Gauss lead to strong deflection
even at the highest energies. This goes along with a time delay
$\tau(E,r)\simeq r\theta(E,d)^2/4$, or
\begin{eqnarray}
\tau(E,r)&\simeq&1.5\times10^3\,{\rm yr}\,Z^2
\left(\frac{E}{10^{20}\,{\rm eV}}\right)^{-2}\label{delay}\\
&&\times\left(\frac{r}{10\,{\rm Mpc}}\right)^{2}
\left(\frac{l_c}{{\rm Mpc}}\right)
\left(\frac{B}{10^{-9}\,{\rm G}}\right)^2\,,\nonumber
\end{eqnarray}
which can be millions of years. A source visible in UHECR today
could therefore be optically invisible since many models involving,
for example, active galaxies as UHECR accelerators, predict
variability on shorter time scales. Strong deflection also
limits the distance $r$ a UHECR of given energy can travel during
its energy loss time, which sometimes is called a ``magnetic
horizon''~\cite{sse,deligny}. For example, if energy losses are
included, for a space-filling field $B\sim1\,$nG with
$l_c=1\,$Mpc the propagation distance of UHECR above $10^{19}\,$eV
is limited to $\sim400\,$Mpc~\cite{sse}. For fields $B\sim0.1\,\mu$G
the magnetic horizon would be only $\sim20\,$Mpc~\cite{deligny}.

\section{Numerical Simulations: Large Scale Structure}
Quite a few simulations of the effect of extragalactic magnetic fields
(EGMF) on UHECR exist in the literature, but usually idealizing
assumptions concerning properties and distributions of sources
or EGMF or both are made: In Refs.~\cite{slb,ils,lsb,sse,is} sources
and EGMF follow a pancake profile mimicking the local supergalactic
plane. In other studies EGMF have been approximated
in a number of fashions: as negligible~\cite{sommers,bdm},
as stochastic with uniform statistical properties~\cite{bo,ynts,ab},
or as organized in spatial cells with a given coherence length and a strength
depending as a power law on the local density~\cite{tanco}.
Recently, Ref.~\cite{mte,sme} have carried out the first attempts to 
simulate UHECR propagation in a realistically structured universe;
additional, independent calculations have now followed
~\cite{dolag}. So far all of these simulations have been limited to
the case of nucleons.

\begin{figure}
\includegraphics[width=0.48\textwidth,clip=true]{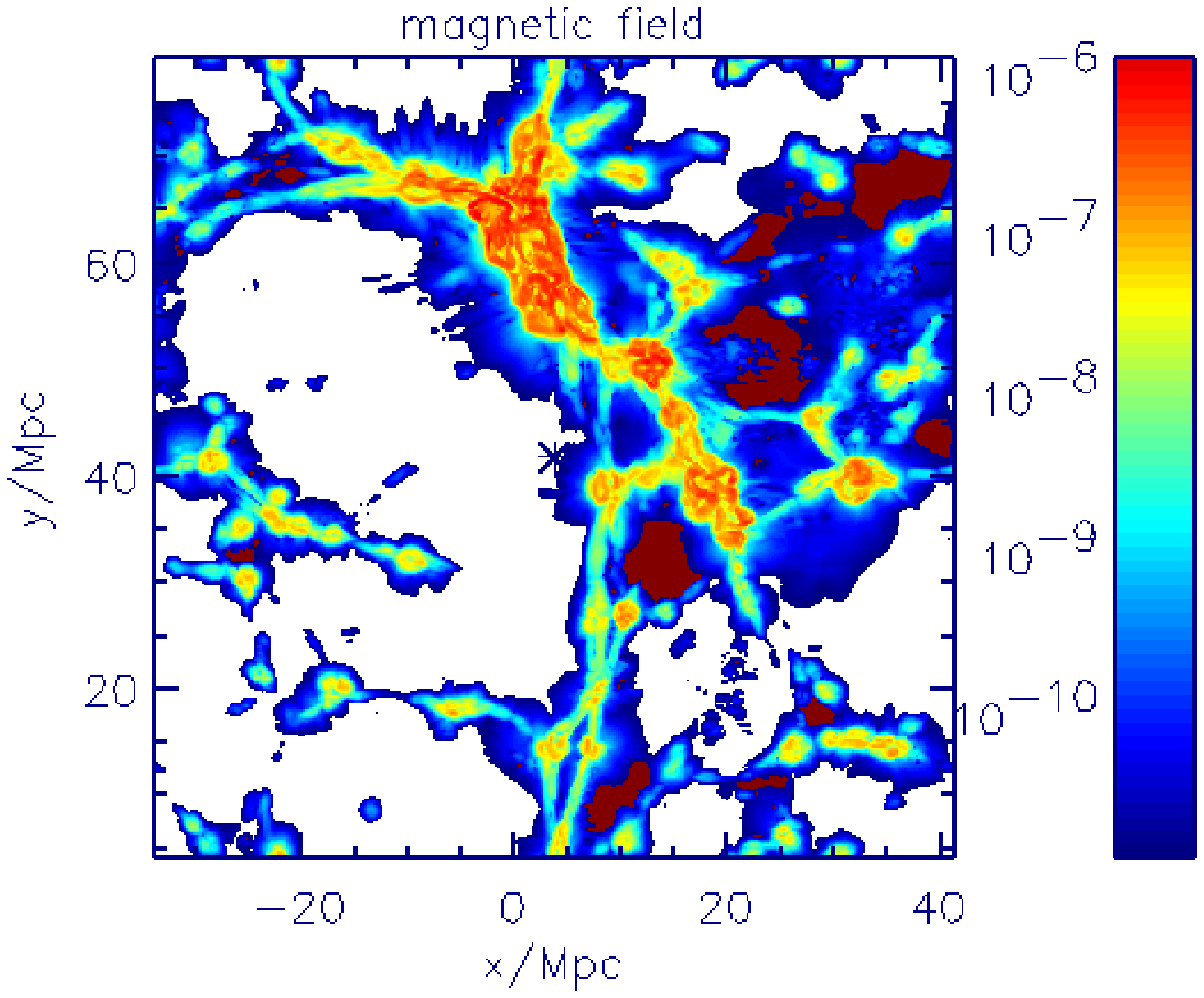}
\includegraphics[width=0.48\textwidth,clip=true]{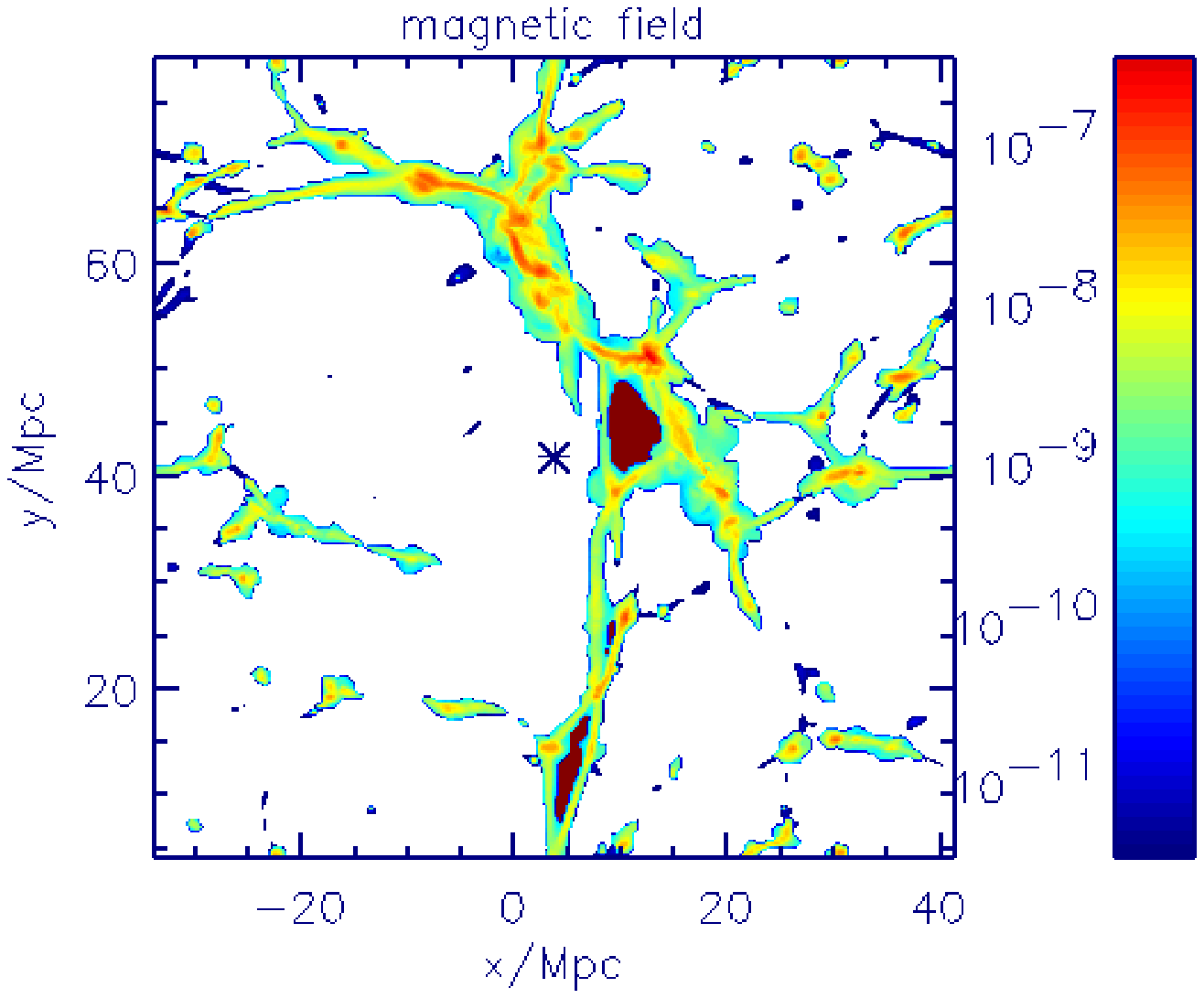}
\caption{Log-scale two-dimensional cuts through
magnetic field total strength in Gauss (color
scale in Gauss) for two EGMF scenarios. In the upper panel
seed fields were injected at shocks, as in the scenario
studied in Ref.~\cite{sme}, whereas in 
the lower panel, an initial uniform seed field was assumed.
In both cases the seeds are chosen such that the final fields
are of order $\sim1\mu\,$Gauss in a Coma-like cluster.
For the latter case, fields in
the voids were suppressed to minimize deflection.
The observer is in the center of the figures and is marked by a star.
The EGMF strength at the observer is $\simeq10^{-11}\,$G in the
upper panel and $\sim0$ in the lower panel. The roughly spherically
magnetized region about 3 Mpc to the lower right of the observer
in the upper panel was used to simulate UHECR fluxes from
an individual source, see Sect.~\ref{sec3}.}
\label{fig1}
\end{figure}

In Ref.~\cite{sme} the magnetized extragalactic environment used for
UHECR propagation is produced by a simulation of the large scale
structure of the Universe. The simulation was carried out within a
computational box of $50\,h^{-1}\,$Mpc length on a side, with
normalized Hubble constant $h\equiv H_0/(100$ km s$^{-1}$ Mpc$^{-1})$
= 0.67, and using a comoving grid of 512$^3$ zones and 256$^3$ dark
matter particles. The EGMF was initialized to zero at simulation start
and subsequently its seeds were generated at cosmic shocks through the
Biermann battery mechanism~\cite{kcor97}. Since cosmic shocks form
primarily around collapsing structures including filaments, the above
approach avoids generating EGMF in cosmic voids.  In this particular
case (more are explored, see below) the resulting magnetic fields in
collapsed structures are rather extended, as can be seen in
Fig.~\ref{fig1} (top panel), significantly more than in the case of a
uniform initial seed field (Fig.~\ref{fig1} lower panel).
The resulting EGMF has been shown to be compatible with
existing Faraday rotation measures with lines of sight both through
clusters and the diffuse intergalactic medium~\cite{ryu}.

In Ref.~\cite{sme} the following question was addressed: which
observer positions, and source distributions and characteristics
lead to UHECR distributions whose spherical multi-poles for $l\leq10$
and auto-correlation at angles $\theta\la20^\circ$ are consistent
with observations ? It was found that (i) the observed
large scale UHECR isotropy requires the neighborhood within a few Mpc
of the observer is characterized by weak magnetic fields below $0.1\,\mu$G,
and (ii) once that choice is made, current data do not strongly
discriminate between uniform and structured source distributions
and between negligible and considerable deflection. Nevertheless,
current data moderately favor a scenario in which (iii) UHECR
sources have a density $n_s\sim10^{-5}\,{\rm Mpc}^{-3}$ and follow the matter
distribution and (iv) magnetic fields are relatively pervasive within the large
scale structure, including filaments, and with a strength of order of a $\mu$G
in galaxy clusters. A two-dimensional cut through the
EGMF environment of the observer in a typical such scenario is
shown in Fig.~\ref{fig1}.

It was also studied in Ref.~\cite{sme} how future data of considerably
increased statistics can be used to learn more about EGMF and source
characteristics. In particular, low auto-correlations at
degree scales imply magnetized sources quite independent of
other source characteristics such as their density. The source
characteristics can only be estimated from the auto-correlations
halfway reliably if magnetic fields have negligible impact on propagation.
Otherwise, the images of sources immersed in considerable magnetic
fields are smeared out, which also smears out the auto-correlation
function over several
degrees. For a sufficiently high source density, individual images
can thus overlap and sensitivity to source density is consequently
lost. The statistics expected from next generation experiments
such as Pierre Auger~\cite{auger} and EUSO~\cite{euso} should
be sufficient to test source magnetization by the auto-correlation
function~\cite{sme}.

\begin{figure}
\includegraphics[width=0.48\textwidth,clip=true]{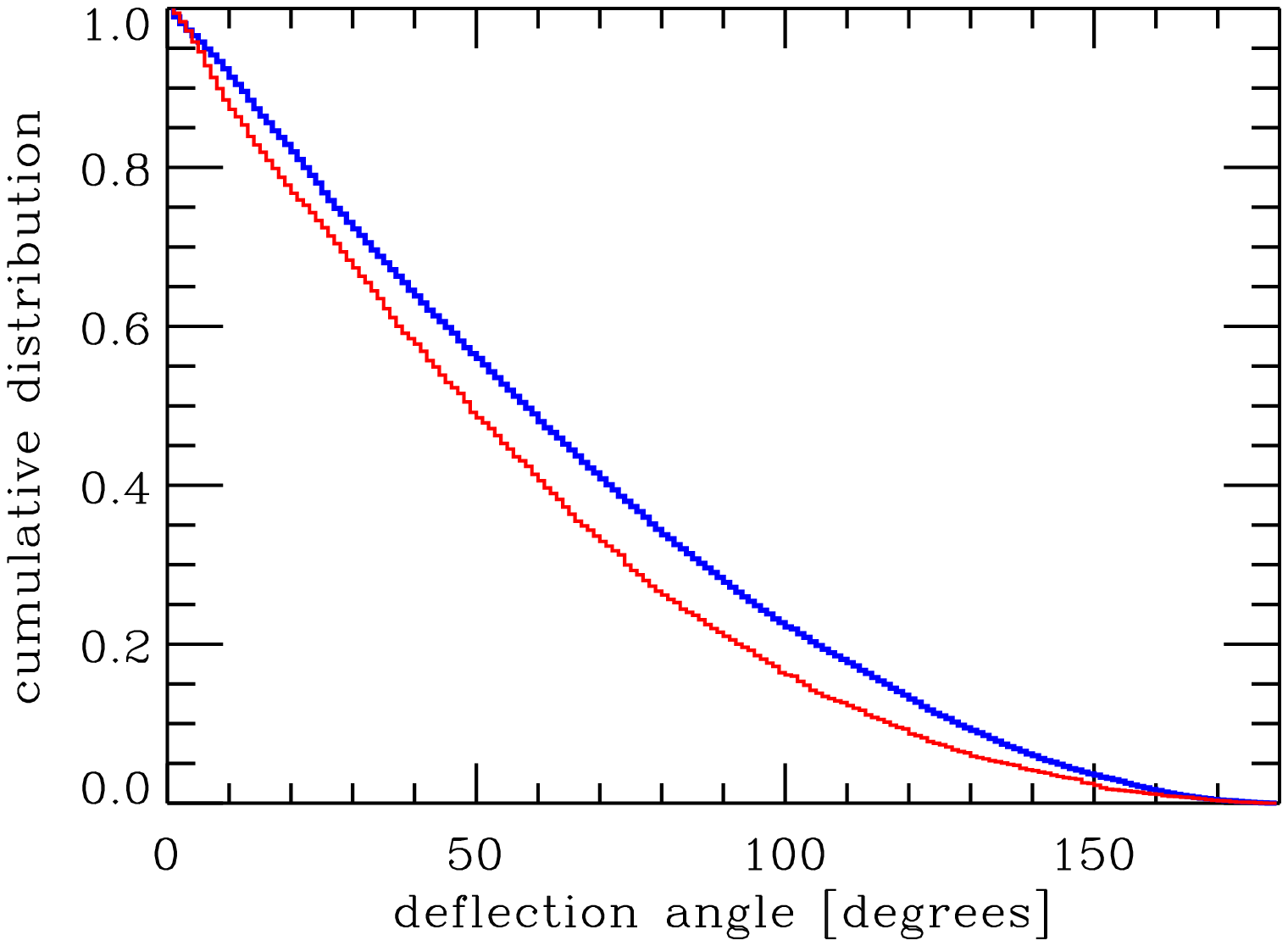}
\includegraphics[width=0.48\textwidth,clip=true]{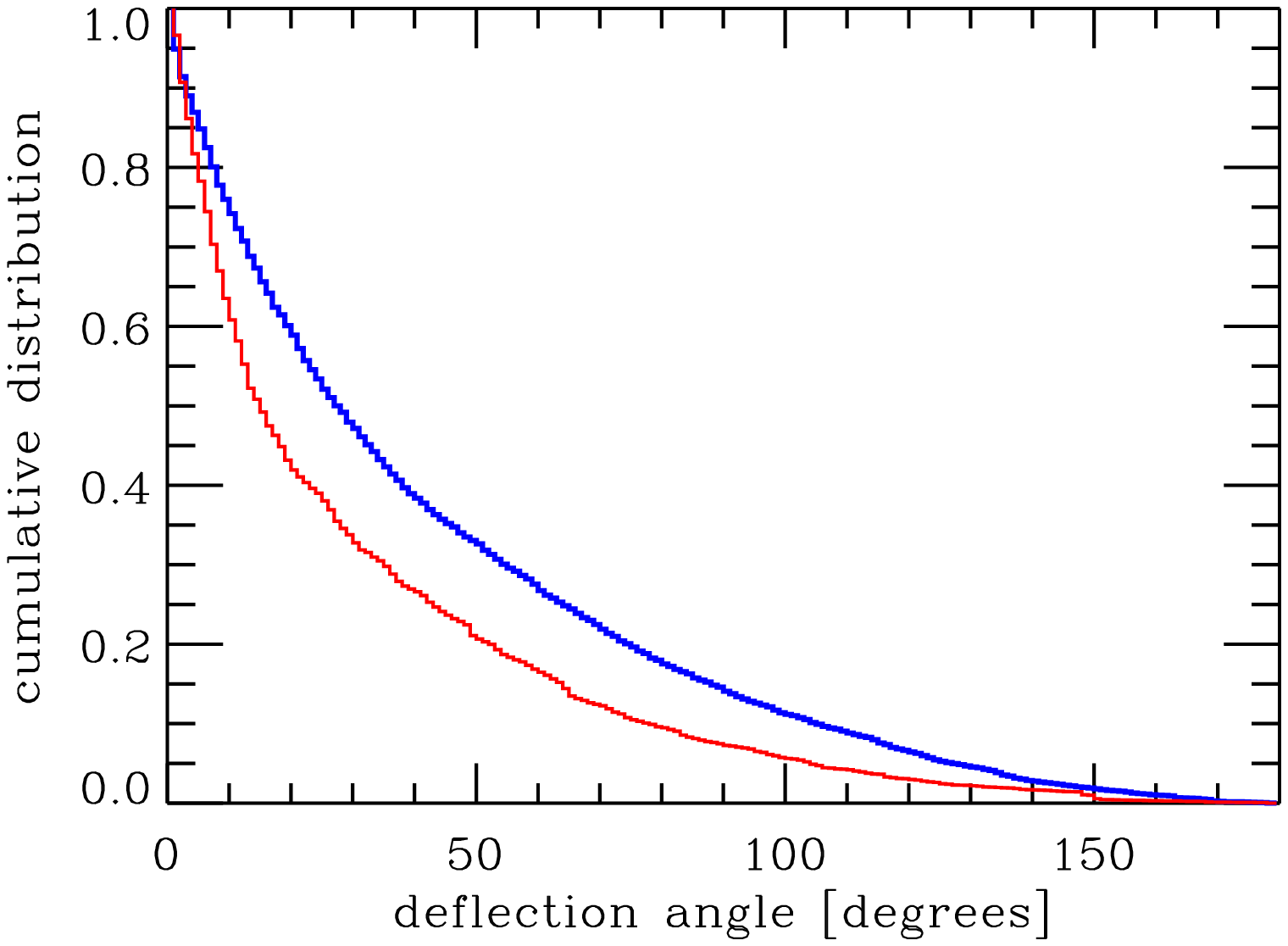}
\caption{The cumulative distribution of deflection
angles $\alpha$ with respect to the line of sight to the sources
of cosmic rays above $4\times10^{19}\,$eV (upper panel) and
above $10^{20}\,$eV (lower panel). The blue (thick) and red (thin)
curves showing larger and smaller deflections, respectively,
are for the two EGMF scenarios from Fig.~\ref{fig1} upper and lower
panel, respectively. The sources follow the baryon density and have average
density $n_s=2.4\times10^{-5}\,{\rm Mpc}^{-3}$.
Shown are the averages over several realizations varying in the positions
and luminosities $Q_i$ of individual sources,
the latter assumed to be distributed as $dn_s/dQ_i\propto Q_i^{-2.2}$
with $1\leq Q_i\leq100$ in arbitrary units.}
\label{fig2}
\end{figure}

In Ref.~\cite{dolag} a constrained simulation which approximately 
reproduces the local universe on scales between a few Mpc and
115 Mpc from Earth was performed and the magnetic smoothed particle
hydrodynamics technique was used to follow the EGMF evolution. The
EGMF was seeded by a uniform seed field of maximal strength compatible
with observed rotation measures in galaxy clusters.

The questions considered in this work were somewhat different. 
In Ref.~\cite{dolag} deflections of UHECR above
$4\times10^{19}\,$eV were computed as a function of the direction
to their source which were assumed to be at cosmological distances.
Specific source distributions were not considered. 
The deflections typically were found to be smaller than a few degrees.

Interestingly, however, there are considerable quantitative differences
in the typical deflection angles predicted by the two EGMF models
in Refs.~\cite{sme,dolag}. These are {\it not} due to
specific source distributions: In fact, for homogeneous source distributions,
the average deflection angle for UHECR above $4\times10^{19}\,$eV
obtained in Ref.~\cite{sme} is $\simeq61^\circ$ above $4\times10^{19}\,$eV
and $\sim33^\circ$ above $10^{20}\,$eV, much larger than in
Ref.~\cite{dolag}.

In addition, even if the magnetic field strength is reduced by a
factor 10 in the simulations in the environment of Fig.~\ref{fig1},
upper panel, the average
deflection angle is still $\sim28^\circ$ above $4\times10^{19}\,$eV,
and $\sim10^\circ$ above $10^{20}\,$eV. This non-linear behavior of
deflection with field normalization is mostly due to the strongly
non-homogeneous character of the EGMF.

Recently we have carried out a new simulation in which the initial
magnetic seeds are provided by a uniform magnetic field (instead of
the Biermann battery) to check how this different model affects our
results.  However, we find that typical deflection angles change at
most by a factor of 2.  This is demonstrated by Fig.~\ref{fig2} which
compares deflection angle distributions for the two EGMF scenarios
shown in Figs.~\ref{fig1} based on the Biermann battery
model or uniform initial seeds, respectively. 
In both the  above cases the
magnetic fields in a Coma-like cluster at $z=0$ were of order
$\mu$G.  Also, in the case of uniform initial fields we have further
suppressed fields in the voids to minimize deflections.

Thus the residual differences in the predictions for the deflection
angles of UHECR are probably due to the different numerical models
for the magnetic fields.  Numerical issues may play an important
role because they affect the amplification and the topological
structure of the magnetic fields, both of which are important for the
normalization procedure, see below.  A few examples are provided in the 
following. The resolution in Ref.~\cite{sme}
is constant and in general better in filaments and voids but worse in the
core of galaxy clusters than the (variable) resolution in
Ref.~\cite{dolag}. Magnetohydrodynamic (MHD) codes are affected
by numerical viscosity to a
much larger extent than hydro codes; for this reason is it
advantageous to evolve the magnetic field as a passive quantity
because, while the amplification due to compression and stretching is
fully accounted for, the additional dissipation introduced by the more
complex MHD scheme is avoided.  This is even more worrisome for the
case of Ref.~\cite{dolag} where the numerical dissipation is not
quantified.  It is also worth noting that the algorithm employed in
Ref.~\cite{dolag} does not guarantee a divergence-free magnetic field.
While it is argued that in the high density regions (core of galaxy
clusters) the variations of the magnetic fields on a scale of a few
resolution elements are larger than the divergence-component, such
conditions do not apply as easily in the low density, low resolution
regions such as in filaments. Thus, one may wonder about the
reliability of the numerical solution there.

In conclusion, while in both simulations the magnetic fields are
normalized to (or reproduce) the same ``observed'' values in the core
of rich clusters, their values in the filaments are substantially
different. While in Ref.~\cite{sme} the amplification in cluster cores
may be underestimated, the magnetic fields in the filaments are not ruled
out by available observational data~\cite{ryu}. On the other hand,
there is some concern for the reliability
of the numerical results of Ref.~\cite{dolag} in the low density regions,
because numerical effects have not been quantified there. 

\begin{figure}
\includegraphics[width=0.48\textwidth,clip=true]{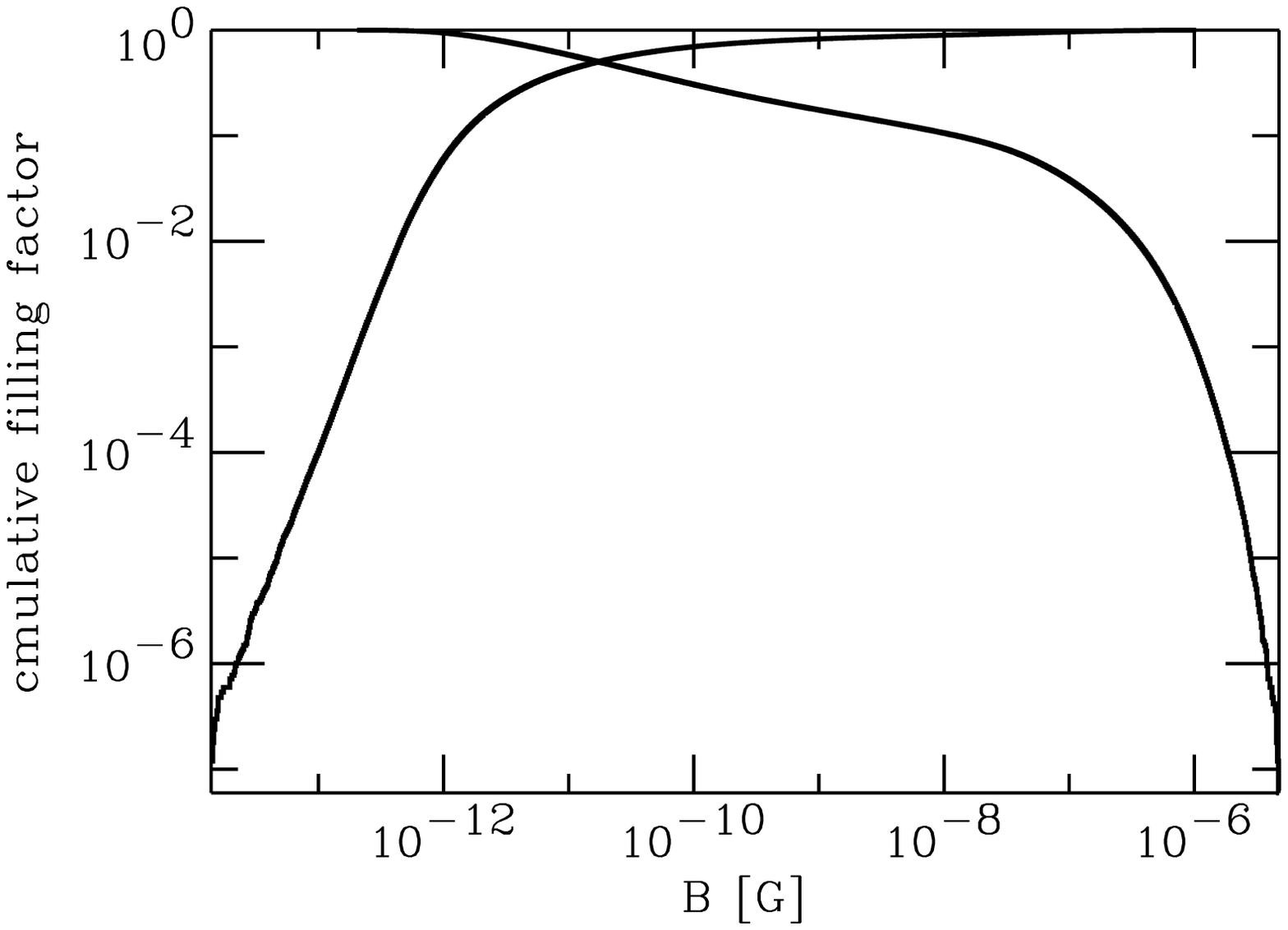}
\includegraphics[width=0.48\textwidth,clip=true]{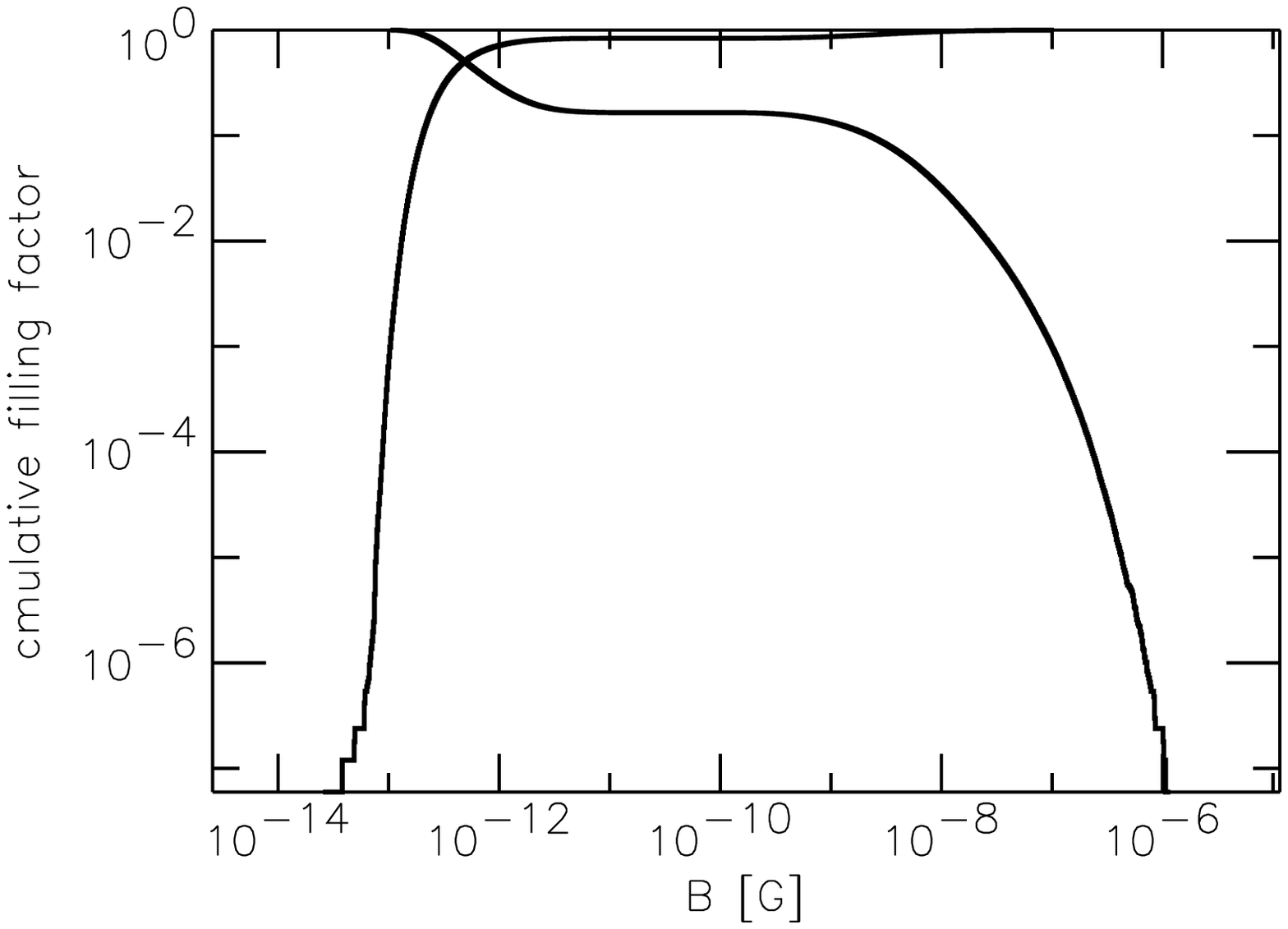}
\caption{The cumulative filling factors for EGMF strength
in the two simulations shown in Fig.~\ref{fig1} above (decreasing curve)
and below (increasing curve) a given threshold, as a function of
that threshold. Upper and lower panels correspond to the panels
in Fig.~\ref{fig1}.}
\label{fig3}
\end{figure}

As a result, the magnetic fields obtained
in Ref.~\cite{sme} are considerably more extended than the ones
in Ref.~\cite{dolag}. This can be seen by comparing
Fig.~\ref{fig3} with a similar plot in Ref.~\cite{dolag_proc}:
For the case shown in the upper panel of Fig.~\ref{fig3},
about 10\% of the volume is filled with fields
stronger than 10 nano Gauss, and a fraction of $10^{-3}$ is
filled by fields above a micro Gauss. Corresponding filling
factors for these field strengths in case of Ref.~\cite{dolag_proc}
are $\la10^{-4}$ and $\la3\times10^{-6}$, respectively. This is
the most likely reason for the much larger average deflections obtained
in Ref.~\cite{sme} as compared to the degree scale deflections
discussed in Ref.~\cite{dolag}.

Since very little is currently known about the
properties of intergalactic magnetic fields, the only solid conclusion
that can be drawn at this stage is that the effects of EGMF on UHECR
propagation are currently rather uncertain.

Finally we note that these studies should be extended to include heavy
nuclei~\cite{prepa} since there are indications that a fraction as
large as 80\% of iron nuclei may exist above
$10^{19}\,$eV~\cite{watson}.  Preliminary results~\cite{prepa} give
deflections of $\sim60^\circ$ above $10^{20}\,$eV in our standard
scenario shown in Figs.~\ref{fig1} and~\ref{fig3}, upper panels. More
importantly, even in the EGMF scenario of Ref.~\cite{dolag},
deflections could be considerable and may not allow particle astronomy
along many lines of sight: The distribution of deflection angles in
Ref.~\cite{dolag} shows that deflections of protons above
$4\times10^{19}\,$eV of $\ga1^\circ$ cover a considerable fraction of
the sky. Suppression of deflection along typical lines of sight by
small filling factors of deflectors is thus unimportant in this case.
The deflection angle of any nucleus at a given energy passing through
such areas will therefore be roughly proportional to its charge as
long as energy loss lengths are larger than a few tens of
Mpc~\cite{bils}. Deflection angles of $\sim20^\circ$ at
$\sim4\times10^{19}\,$eV should thus be the rule for iron nuclei. In
contrast to the contribution of our Galaxy to deflection which can be
of comparable size but may be corrected for within sufficiently
detailed models of the galactic field, the extra-galactic contribution
would be stochastic. Statistical methods are therefore likely to be
necessary to learn about UHECR source distributions and
characteristics. In addition, should a substantial heavy composition
be experimentally confirmed up to the highest energies, some sources
would have to be surprisingly nearby, within a few Mpc, otherwise only
low mass spallation products would survive propagation~\cite{er}.

The putative clustered component of the UHECR flux whose fraction of the
total flux seems to increase with energy~\cite{teshima1} may play a key role
in this context. It could be caused by discrete sources in directions
with small deflection. Since, apart from energy losses, cosmic rays
of same rigidity $Z/A$ are deflected similarly by cosmic magnetic
fields, one may expect that the composition of the clustered component
may become heavier with increasing energy. Indeed, in Ref.~\cite{teshima}
it was speculated that the AGASA clusters may be consistent with
consecutive He, Be-Mg, and Fe bumps.

\section{Numerical Simulations: Discrete Sources}\label{sec3}
\begin{figure}
\includegraphics[width=0.5\textwidth,clip=true]{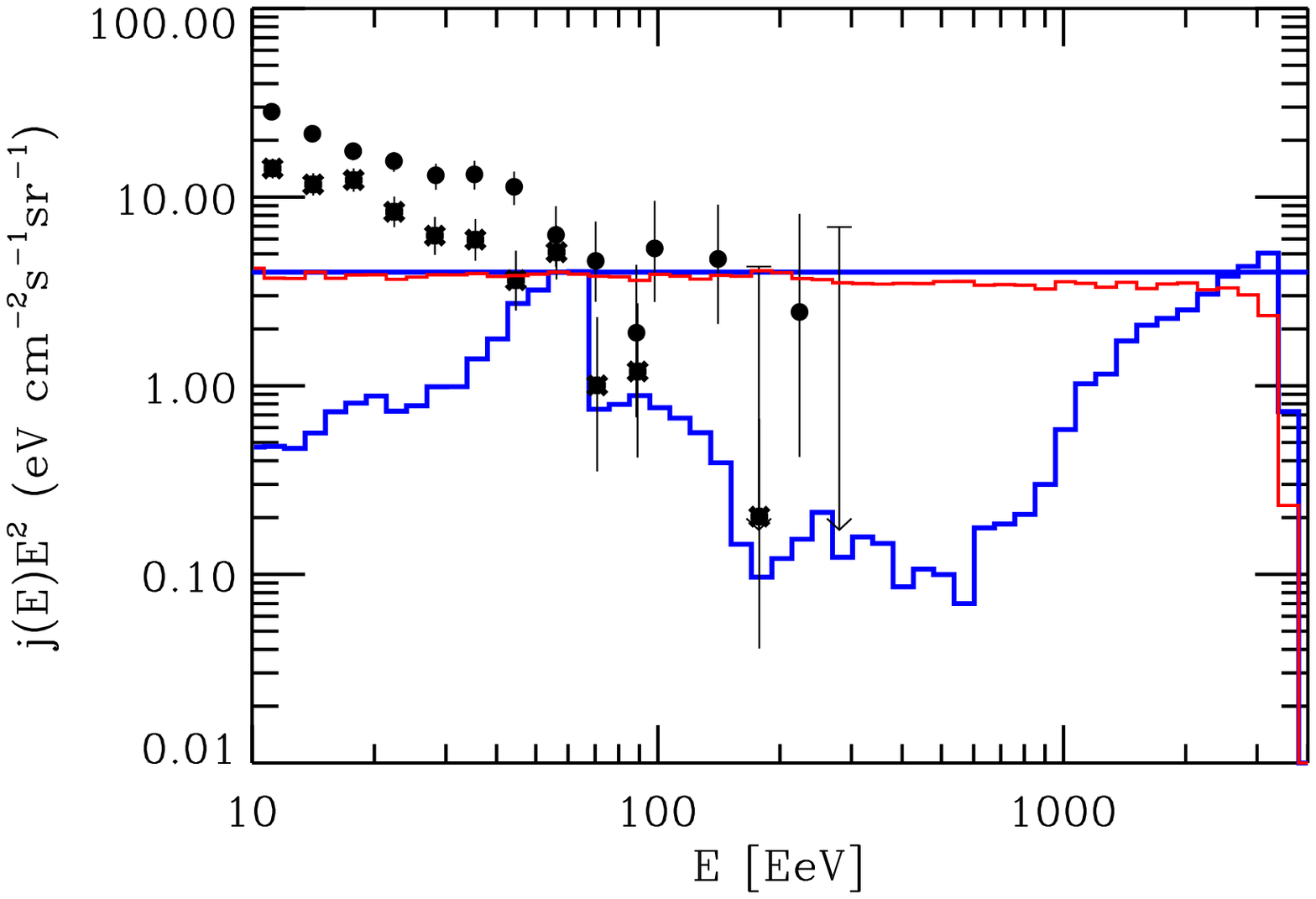}
\includegraphics[width=0.5\textwidth,clip=true]{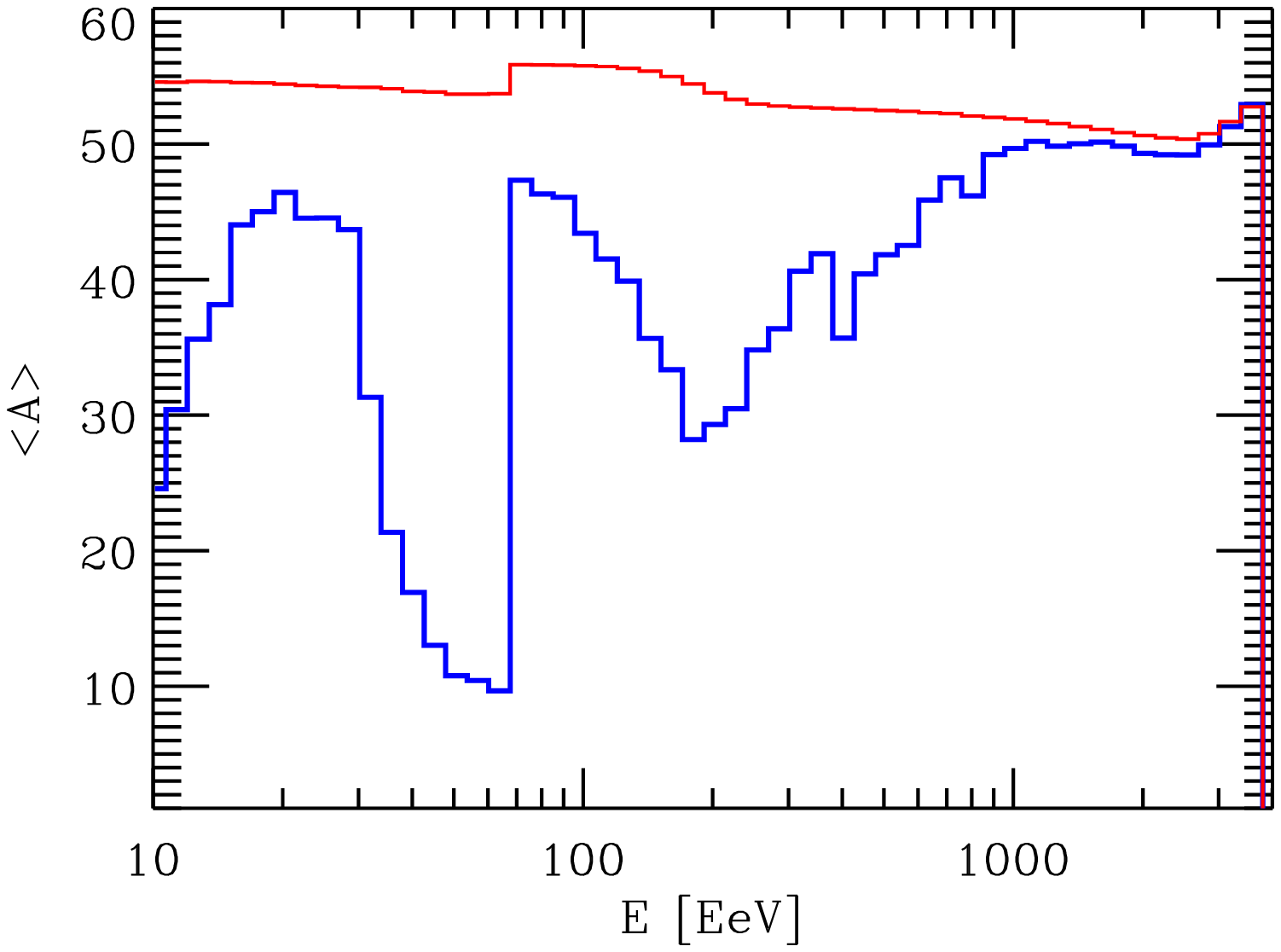}
\caption[...]{Upper panel: Steady state all-particle spectra predicted by a
scenario with a source at 3.3 Mpc distance from Earth injecting
iron primaries with a spectrum $\propto E^{-2.0}$ up to
$4\times10^{21}\,$eV. The blue (thick) curves are for an
EGMF of strength $\sim10^{-7}\,$G surrounding the source
and the red (thin) curve is without EGMF.
Shown for comparison are the solid angle integrated
AGASA~\cite{agasa} (dots) and HiRes-I~\cite{hires} (stars) data.
The solid straight line marks the injection spectrum. All fluxes
have been normalized at 60 EeV. Lower
panel: The predicted composition in the same scenario.}
\label{fig4}
\end{figure}

This brings us to the impact of EGMF on UHECR fluxes from discrete
sources. In Ref.~\cite{sigl} we investigated the impact on cosmic
ray observations above $10^{19}\,$eV of Mpc-scale magnetic fields
of $\sim10^{-7}\,$G strength surrounding UHECR sources. The likely
case was assumed that magnetic fields
within a few Mpc around Earth are insignificant. The trajectory
simulations are based on the same large scale structure simulation
as before, examples of whose EGMF distributions were shown
in Figs.~\ref{fig1} and~\ref{fig3}, upper panels. We find that
such source fields can strongly modify spectra and composition
at Earth, especially for nearby sources for which the fields can
considerably increase propagation times relative to both energy
loss and photo-disintegration time scales and to the undeflected
propagation time: Eq.~(\ref{delay}) shows qualitatively that time
delays at $E\sim10^{20}\,$eV can easily reach $\sim10^7\,$years
for fields $B\sim10^{-7}\,$G extended over a few Mpc. This is indeed
larger than energy loss times $\sim3\times10^6\,$years and comparable
to the straight line propagation time $\sim10^7\,$years.
We found the following generic features:

We note in passing that spallation and pion production could
be enhanced around powerful sources due to an increased
infra-red background~\cite{wolfendale}. While we have not yet taken this
into account in the present simulations, we expect this effect
to be small compared to the EGMF effect.

The spectra are considerably
hardened relative to the injection spectrum at energies below
the usual GZK-like cutoff where energy loss distance and source
distance become comparable. This is caused by an interplay between
diffusion and energy loss: The flux of low energy particles is
suppressed because diffusion spreads them out over a larger volume
due to their much larger energy loss times. This is in contrast
to the case of uniformly distributed magnetic fields which in
general lead to a steepening of the cosmic ray flux below the
GZK cutoff. A hardened sub-GZK spectrum from individual sources
would indeed be consistent with the hints of a hard clustered component in
the AGASA data between $10^{19}\,$eV and $10^{20}\,$eV~\cite{teshima1}.

Furthermore, for a nucleus of atomic mass $A$ as injected primary, due
to the kinematics of the photo-disintegration reactions a nucleon
peak appears at energy $\sim E_{\rm max}/A$, where $E_{\rm max}$ is
the maximal nucleus injection energy. This effect is the more prominent
the harder the injection spectrum.
We also found that the details of spectra and composition
depend significantly on the unknown details of the magnetic
fields and the position of the source therein and can thus
not be predicted.

An example demonstrating these effects is shown in Fig.~\ref{fig4}
where a source of iron nuclei at 3.3 Mpc
distance to Earth is considered, once without any EGMF, and
once with an EGMF concentrated around the source and reaching
$\sim10^{-7}\,$G there. The observer is at the same position
as indicated in Fig.~\ref{fig1} and the source is in the center of the
roughly spherically magnetized region to the lower right of the
observer in the upper panel of Fig.~\ref{fig1}. The spectral
modification for proton primaries (not shown) by the EGMF would
be equally severe as in Fig.~\ref{fig4}~\cite{sigl}.

Further cutoffs towards low energies can be induced if
the source is active only since a time smaller than the
typical delay time at this energy. Since the latter easily
reaches $\sim10^9\,$years at $10^{19}\,$eV in the scenario shown
in Fig.~\ref{fig4}, as can also be seen from Eq.~(\ref{delay}),
this is actually quite likely for active galactic nuclei which
can have activity time scales below $\sim10^8\,$years.
Obviously, the characteristics of
such features also depend on unknown details of the source.

Our simulations finally show that even for iron primaries,
extra-galactic magnetic fields from large scale structure
simulations are not strong and extended enough to explain the
observed large scale isotropy of UHECR
arrival directions in terms of a single nearby source. This
would require more homogeneous fields such as in Ref.~\cite{agrs}.

\begin{figure}[ht]
\includegraphics[width=0.5\textwidth,clip=true]{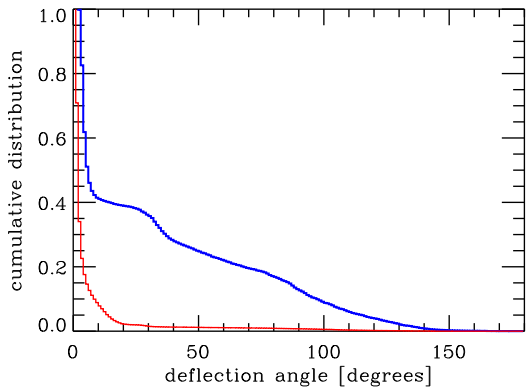}
\caption[...]{The cumulative distribution of arrival direction
off-sets from the source direction for UHECR above $4\times10^{19}\,$eV
in the scenario described in the text for iron primaries, corresponding to
Fig.~\ref{fig4} (blue, thick curve) and proton primaries (red, thin
curve).}
\label{fig5}
\end{figure}

Next generation experiments such as the Pierre Auger
Observatories~\cite{auger} and the EUSO project~\cite{euso}
will accumulate sufficient statistics to establish
spectra and distributions of composition and arrival directions
from individual sources. A potentially strong influence of magnetic
fields surrounding individual sources should thus be kept in mind when
interpreting data from these experiments. This is true even
if UHECR arrive within a few degrees from the source position.
As Fig.~\ref{fig5} shows~\cite{sigl}, this is in fact the case for proton
primaries in our example for a discrete source.

\section{Conclusions}
It thus seems evident that the influence of large
scale cosmic magnetic fields on ultra-high energy cosmic ray propagation
is currently hard to quantify and may not allow to do ``particle
astronomy'' along most lines of sight, especially if a significant
heavy nucleus component is present above $10^{19}\,$eV. However,
even in our simulations, there are lines of sight along which
deflection is only a few degrees, at least for proton primaries.
Such directions might still be suitable for ``particle astronomy''.

\section*{Acknowledgments}
GS thanks Antonio Insolia for inviting me to the CRIS meeting
and for financial support.

\end{document}